\documentclass[10pt,a4paper]{article}

\usepackage[left=2cm,right=2cm,top=2cm,bottom=2cm]{geometry}
\usepackage[english]{babel}
\usepackage{amsmath}
\usepackage{amsfonts}
\usepackage{amssymb}
\usepackage{bbm}
\usepackage{latexsym}
\usepackage{lmodern}
\usepackage{mathrsfs}
\usepackage{slashed}
\usepackage{braket}
\usepackage{tensor}
\usepackage{simplewick}
\usepackage{graphicx}
\usepackage{xcolor}
\usepackage[font=small,labelfont=it]{caption}
\usepackage[font=scriptsize]{caption}
\usepackage[hidelinks]{hyperref}
\hypersetup{
        colorlinks=true,
        linkcolor=blue,
        citecolor=magenta,
}
\usepackage[nottoc]{tocbibind} 

\def\C{\mathbb C}
\def\R{\mathbb R}
\def\NN{\mathbb N}

\newcommand{\sfrac}[2]{{\textstyle\frac{#1}{#2}}}

\newcommand{\pa}{\partial}
\newcommand{\im}{{\mathrm{i}}}

\newcommand{\ep}{{\mathrm{e}}}
\newcommand{\diff}{{\mathrm{d}}}

\newcommand{\beq}{\begin{equation}}
\newcommand{\eeq}{\end{equation}}
\newcommand{\eq}{\end{equation}}
\newcommand{\bea}{\begin{eqnarray}}
\newcommand{\eea}{\end{eqnarray}}

\renewcommand{\and}{{\quad{\rm and}\quad}}
\newcommand{\und}{{\qquad{\rm and}\qquad}}
\renewcommand{\=}{\ =\ }

\newcommand{\nn}{\nonumber}

\allowdisplaybreaks  
\DeclareMathOperator{\tr}{tr}

\DeclareMathOperator{\Tr}{Tr}

\newcommand{\Lagr}{ { \mathcal{L} } }
\newcommand{\Delt}{ { \mathcal{M} } }
\newcommand{\Lagrc}{\ensuremath{\mathring{\mathcal{L}}}}
\newcommand{\Deltc}{\ensuremath{\mathring{\mathcal{M}}}}

\advance \textheight by \topskip
\makeatletter
\@addtoreset{equation}{section}

\makeatother
\usepackage{tikz}
\usetikzlibrary{positioning,arrows}
\usetikzlibrary{decorations.pathmorphing}
\usetikzlibrary{decorations.markings}
\tikzset{
particle/.style={thick,draw=black, postaction={decorate},
    decoration={markings,mark=at position .50 with {\arrow[line width=1pt]{>}}  }},
aparticle/.style={thick,draw=black, postaction={decorate},
    decoration={markings,mark=at position .5 with  {\arrow[line width=1pt]{<}}  }},
gluon/.style={decorate, draw=black,
    decoration={coil,aspect=0}},
scalar/.style={thick,dashed,draw=black}
 }

\begin{document}

\begin{titlepage}
\setcounter{page}{0}

\phantom{.}
\vskip 1.5cm

\begin{center}

{\LARGE \bf 
Nicolai maps for supersymmetric sigma models
}
\vspace{12mm}

{\Large Olaf Lechtenfeld}
\\[8mm]
\noindent {\em
Institut f\"ur Theoretische Physik and Riemann Center for Geometry and Physics\\
Leibniz Universit\"at Hannover, Appelstrasse 2, 30167 Hannover, Germany}
\vspace{12mm}

\begin{abstract} 
\noindent
Supersymmetric field theories can be characterized by their Nicolai map,
which is a nonlinear and nonlocal field transformation to their free-field limit.
The systematic construction of such maps has recently been outlined 
for actions with power more than two in the fermions, which produces
a perturbative expansion in loop-decorated fermionic tree diagrams.
We thoroughly investigate the nonlinear $\mathbb C P^1$ sigma model 
in ($3{+}1$)-dimensional Minkowski space as a paradigmatical example.
We construct and test a chiral form of the Nicolai map, to third order 
in the coupling, including all (regularized) quantum parts.
In addition, all trees with one or two edges are summed up.
The free-action condition determines only the one-edge part of the map.
We resolve the fermion loop decoration of the Nicolai trees by injecting an 
auxiliary vector field and present the ensuing classical Nicolai map to 
second order in a dimensionful coupling.
\end{abstract}

\end{center}

\end{titlepage}
%


\section{Introduction and summary}

\noindent
When a field theory is supersymmetric then it enjoys a (generically but not always)
nonlocal and nonlinear field transformation relating it at different values of its parameters,
say coupling constants~$g$. This so-called Nicolai map~$T$~\cite{Nic1,Nic2,Nic3}
connects the quantum expectation value of any operator~$Y$ built from the bosonic fields~$\phi$
at different coupling values,
\begin{equation} \label{Tdef}
\bigl\langle\!\!\!\bigl\langle\!\!\!\bigl\langle Y[\phi] \bigr\rangle\!\!\!\bigr\rangle\!\!\!\bigr\rangle_g
\= \bigl\langle\!\!\!\bigl\langle\!\!\!\bigl\langle (T_{gg'}^{-1}Y)[\phi] \bigr\rangle\!\!\!\bigr\rangle\!\!\!\bigr\rangle_{g'}
\= \bigl\langle\!\!\!\bigl\langle\!\!\!\bigl\langle Y[T_{gg'}^{-1}\phi] \bigr\rangle\!\!\!\bigr\rangle\!\!\!\bigr\rangle_{g'}\ .
\end{equation}
Most importantly, it allows one to evaluate such correlators in the free theory (at $g'{=}0$),
which we take as the reference coupling from now on.\footnote{
The zero vacuum energy implied by unbroken supersymmetry normalizes $\langle1\rangle_g=1$.}
The subscript on the correlator and also on the symbol of the map~$T_g:\phi\mapsto \phi'[g,\phi]=T_g\phi$
indicates the value of the coupling.
As spelled out in~\eqref{Tdef}, the Nicolai map is distributive, 
i.e.~$T(\phi_1\phi_2)=(T\phi_1) \, (T\phi_2)$, 
It operates not in the original supersymmetric theory but in a nonlocal bosonic theory
obtained by integrating out all anticommuting degrees of freedom~$\psi$ (and possibly auxiliary fields~$F$).
Hence, the fat-bracket expectation values of~\eqref{Tdef} denote functional averaging over the remaining
dynamical bosonic fields in this effective theory, ruled by an action
\begin{equation} \label{Sloops}
S_g[\phi] \= S_g^{(0)}[\phi]\ +\ \sum_{r=1}^\infty \hbar^r\,S_g^{(r)}[\phi]\ ,
\end{equation}
where the classical local part $S^{(0)}_g$ is the bosonic piece of the original supersymmetric action
$\ensuremath{\mathring{S}}_{\textrm{\tiny SUSY}}[\phi,\psi,F]$ after eliminating auxiliaries, 
and the nonlocal quantum corrections arise from the path integral over the anticommuting fields 
in the partition function, all at coupling~$g$. 
The power~$r$ of $\hbar$ gives the number of fermion loops in the diagrammatic expansion of~$S_g^{(r)}$.

Until recently, only supersymmetric actions quadratic in the fermionic fields were known to 
feature a Nicolai map. In such a case, the loop expansion~\eqref{Sloops} stops at $O(\hbar)$
since only fermion one-loop diagrams appear in the effective action~$S_g$. 
In~\cite{CLR} however, the generalization to arbitrary supersymmetric theories and, hence,
to all fermion loop orders has been achieved. 
A first application has been to super Yang--Mills theory in the light-cone gauge~\cite{L3}.
There is a price to be paid. Beyond quadratic fermions, 
the Nicolai map is no longer classical but also a power series in~$\hbar$,
\begin{equation} \label{quantmap}
T_g\phi \= T_g^{(0)}\!\phi \ +\ \sum_{r=1}^\infty \hbar^r\,T_g^{(r)}\!\phi\ ,
\end{equation}
on top of the formal expansion in powers of~$g$. Also here,
$r$ counts the number of fermion loops in the diagrammatic representation.
Changing path-integral variables $Y\mapsto T_gY$ on the right-hand side of~\eqref{Tdef} and 
comparing the path-integral measures, one finds the identity
\begin{equation} \label{matchingnew}
S^{(0)}_0[T_g\phi]\ +\ \sum_{r\ge1}\hbar^r\,S^{(r)}_0
\ -\ \im\hbar\,\Tr\ln\sfrac{\delta T_g\phi}{\delta\phi}
\= S^{(0)}_g[\phi]\ +\ \sum_{r\ge1} \hbar^r\,S_g^{(r)}[\phi]
\end{equation}
where $\Tr$ refers to the functional trace.
For $g{=}0$ the terms in the sum do not depend on~$\phi$, 
and thus the left-hand sum is a constant.
Inserting~\eqref{quantmap} into~\eqref{matchingnew} and matching powers in~$\hbar$
one obtains an infinite hierarchy of `Nicolai-map conditions', one for each loop level. 
The tree-level and one-loop contributions read
\begin{equation} \label{matching01}
S^{(0)}_0[T_g^{(0)}\!\phi] \= S^{(0)}_g[\phi] \quad\und\quad
S^{(0)}_0[T_g\phi]\big|_{O(\hbar)} + S^{(1)}_0 
- \im\,\Tr\ln\sfrac{\delta T_g^{(0)}\!\phi}{\delta\phi} \= S_g^{(1)}[\phi] \ .
\end{equation}
The first relation is known as the `free-action condition' and involves only the classical map,
while the second identity differs from the former `determinant-matching condition'
by a contribution of~$T_g^{(1)}\phi$.\footnote{
Without this term, the left-hand side comes from the Jacobian determinant of the classical
Nicolai map, and the right-hand side arises from the fermion determinant
(for supersymmetric actions quadratic in the fermions).}

There exists a formalism and a universal formula which yields a formal power series expansion 
(in~$g$ and~$\hbar$) of the map and  its inverse~\cite{FL,DL1,L1,L2,ALMNPP,LR1,LR3,Lprague}.\footnote{
Under certain conditions the construction works also in the absence of off-shell supersymmetry,
e.g.~for super Yang--Mills theory in dimensions 6 and~10 in the Landau gauge~\cite{ALMNPP,MN,LR2,LR4}.}
Its main ingredient is the so-called `coupling flow operator' 
\begin{equation}
R_g[\phi] \= \int\!\diff x\ \bigl(\pa_g T_g^{-1} \circ T_g \bigr) \phi(x)\,\frac{\delta}{\delta\phi(x)}\ ,
\end{equation}
where `$x$' represents all coordinates the fields depend on.
The infinitesimal Nicolai map is ruled by this functional differential operator,
\begin{equation} \label{localflow}
\pa_g \bigl\langle\!\!\!\bigl\langle\!\!\!\bigl\langle Y[\phi] \bigr\rangle\!\!\!\bigr\rangle\!\!\!\bigr\rangle_g 
\= \bigl\langle\!\!\!\bigl\langle\!\!\!\bigl\langle \bigl( \pa_g + R_g[\phi] \bigr) Y[\phi] \bigr\rangle\!\!\!\bigr\rangle\!\!\!\bigr\rangle_g\ ,
\end{equation}
also acting in the effective bosonic theory.
To construct the coupling flow operator, one observes that, in chiral theories with off-shell supersymmetry,
\begin{equation} \label{defDelta}
\ensuremath{\mathring{S}}_{\textrm{\tiny SUSY}}[\phi,\psi,F] \= 
\int\!\diff x\ \delta_\alpha \ensuremath{\mathring{\Delt}}_\alpha[\phi,\psi,F](x)
\qquad\Rightarrow\qquad
\pa_g \ensuremath{\mathring{S}}_{\textrm{\tiny SUSY}}[\phi,\psi,F] \= 
\int\!\diff x\ \delta_\alpha \pa_g\ensuremath{\mathring{\Delt}}_\alpha[\phi,\psi,F](x) \ ,
\end{equation}
where $\alpha$ is a spinor index (we shall be more concrete later) and the anticommuting 
$\ensuremath{\mathring{\Delt}}_\alpha$ is the penultimate component in the superfield expansion of the superspace action.
Super Yang--Mills theories are more involved because of gauge-fixing complications and often 
show $[\pa_g,\delta_\alpha]\neq0$, so we exclude them here.
Employing~(\ref{defDelta}) and the supersymmetric Ward identity and integrating out the anticommuting 
and auxiliary variables one obtains the coupling flow operator as
\begin{equation} \label{flowop}
R_g[\phi] \= \frac{\im}{\hbar}\int\!\diff x\!\int\!\diff y\ 
\bigl\langle\!\bigl\langle \pa_g\ensuremath{\mathring{\Delt}}_\alpha[\phi](y)\ 
\delta_\alpha \phi(x) \bigr\rangle\!\bigr\rangle\ \frac{\delta}{\delta\phi(x)}\ ,
\end{equation}
where the double bracket $\langle\!\langle\ldots\rangle\!\rangle$ denotes a functional average over the fermionic
as well as over all auxiliary fields in the supersymmetric theory.
If the auxiliary fields occur only linearly under the double bracket,
then we may insert their on-shell values directly into this expression.
Sometimes, a fraction of the supersymmetry suffices in the construction,
leading to some freedom in the spinor-index sum in~\eqref{flowop}.
This can then be exploited to choose a simpler map.
We shall focus on such  a `chiral map' in this work.

The universal formula~\cite{LR1} directly yields the Nicolai map as the $g$-ordered exponential 
of ${-}\!\!\int_0^g\diff{g'}\,R_{g'}$. To evaluate this expression perturbatively,
the $R_g$ action has to be iterated, $R_{g_s\cdots}R_{g_2} R_{g_1}\phi$.
For a $k$-fermion self-interaction in~$\ensuremath{\mathring{S}}_\textrm{SUSY}$,
this process grafts full fermionic $k$-point functions onto previously generated diagrams.
For the Wess--Zumino model ($k{=}2$) it produces only fermionic tree diagrams, 
dressed with bosonic `leaves', hence a classical Nicolai map.
For nonlinear sigma models ($k{=}4$), however, the graphical expansion of the Nicolai map
involves a quartic fermion self-interaction and thus will feature fermionic
trees with all kinds of fermion loops embedded. 

The present paper demonstrates these features for the four-dimensional supersymmetric $\C P^1$ model. 
In the previous work~\cite{CLR} this was initiated (for general~$\C P^N$ models),
but the Nicolai map had been computed (and checked) only to first order in~$g$ and~$\hbar$.
Here, we firstly push the computation to order~$g^3$, classical and quantum, and even to all orders in~$g$ 
for the trees with one or two edges. We see the first branched tree appearing at~$O(g^3)$.
Secondly, it is demonstrated that the free-action condition only determines the `one-edge' part of the
Nicolai map, the almost local contribution due to a single action of $R_g$ on~$\phi$.
A nontrivial spin structure obstructs the collapse of the Nicolai map to single-edged tree diagrams,
so that all possible tree diagrams occur in its perturbative expansion.
Thirdly, we verify that the higher-edge tree contributions to~$T_g^{(0)}\phi$ can only be fixed 
by higher~$\hbar$ powers in~\eqref{matchingnew}, together with the quantum parts~$T_g^{(r>0)}\phi$.
Fourthly, we introduce an auxiliary vector field~$A$ to resolve the fermion loops in the Nicolai map,
exploiting a gauged formulation of the sigma model where fermions appear only quadratically in the action,
thus allowing for a purely classical Nicolai map in $(\phi,A)$ space.
Compared to~\cite{CLR} we expand in a dimensionful coupling $\lambda\sim\sqrt{g}$ and rescale~$A$
to mass dimension~2 so that the free-field limit is regular. This enhanced Nicolai map
is presented to order~$\lambda^2$ and passes the free-action test.

There are several  ways in which the work presented here can be further expanded or generalized.  
Obviously, one may push the perturbative expansion of the Nicolai map to higher orders in~$g$ 
and/or~$\hbar$, or consider other target spaces.
An extension of the formalism with quartic fermions to gauge theories has been initiated~\cite{L3}.
A particularly ambitious goal is  the formulation of a Nicolai map for supergravity, 
possibly elucidating its UV properties.
In the following, we mostly suppress $\hbar$ by putting it to one.

\section{The flow operator}

\noindent
Although our considerations apply to any supersymmetric nonlinear sigma model in 
($3{+}1$)-dimensional Minkowski space~\cite{bertolini},
for simplicity we restrict the analysis to the simplest example, the $\C P^1$ or $O(3)$ model.
It is conveniently constructed from the K\"ahler potential
\begin{equation} \label{K}
K(\Phi^\dagger,\Phi) \=  \frac{\mu^2}{g}\,\log\Bigl[ 1 + \frac{g}{\mu^2}\,\Phi^\dagger\Phi \Bigr]
\= \Phi^\dagger\Phi\ -\ \tfrac12 \tfrac{g}{\mu^2}\bigl(\Phi^\dagger\Phi \bigr)^2 
\ +\tfrac13 \tfrac{g^2}{\mu^4}\bigl(\Phi^\dagger\Phi \bigr)^3\ +\ O(g^3)
\end{equation}
for a chiral superfield (in Wess--Bagger notation~\cite{wessbagger} and $x$~coordinates)
\begin{equation}
\Phi\= \phi
+ \im \theta \sigma^m \bar\theta\, \partial_m \phi
+ \tfrac14 \theta^2 \bar \theta^2\,\Box \phi
+ \sqrt 2 \theta\, \psi
- \tfrac{\im}{\sqrt 2}  \theta^2 \partial_m \psi \sigma^m \bar\theta
+ \theta^2 F
\end{equation}
and its hermitian conjugate $\Phi^\dagger$ in a supersymmetric $D$-term action~\footnote{
An $F$-term with a superpotential~$W$ may be added but is not relevant for our purposes.}
\begin{equation} \label{Kaction}
\ensuremath{\mathring{S}}_{\textrm{\tiny SUSY}} \ = \int\!\diff^4x \ \Lagrc \qquad\textrm{with}\qquad
\Lagrc \= \int\!\diff^2\theta\,\diff^2\bar\theta \ K(\Phi^\dagger ,\Phi)\ .
\end{equation}
Weyl-spinor indices $\alpha,\dot\alpha=1,2$ are suppressed.
The complex bosonic field~$\phi$ parametrizes the 2-sphere target manifold of size~$\sqrt{\mu^2/g}$.
The mass parameter~$\mu$ accounts for the dimensionality of~$\phi$,
and the dimensionless coupling $g$ is introduced for later convenience,
in order to relate to the free theory (on the ``infinite sphere''~$\R^2$).

Performing the superspace integrations in \eqref{Kaction} one finds the component lagrangian
\begin{equation} \label{LagrCP1c}
\begin{aligned}
\Lagrc &\= -f_g^2\,\partial_m \phi\,\partial^m \phi^* 
-\tfrac{\im}{2}f_g^2 \bigl(\psi\,\slashed{\pa}  \bar \psi+\bar\psi\,\bar{\slashed{\pa}} \psi\bigr)
+f_g^2\,F\,F^*
+\im\,\tfrac{g}{\mu^2} f_g^3\,\psi\,(\phi\,\slashed{\pa}\phi^*{-}\phi^*\slashed{\pa}\phi)\,\bar\psi \\
&\qquad\! + \tfrac{g}{\mu^2} f_g^3\,\bigl(\phi\,F\,(\bar\psi\bar\psi)+\phi^*F^*(\psi\psi)\bigr)
-\tfrac12\tfrac{g}{\mu^2} f_g^4 (1{-}2\tfrac{g}{\mu^2}\phi^*\phi)\,(\bar\psi \bar\psi )\,(\psi \psi)
\end{aligned}
\end{equation}
with the abbreviation
\begin{equation}
f_g\ :=\ (1+\tfrac{g}{\mu^2}\,\phi^*\phi)^{-1}
\= 1\ -\ \tfrac{g}{\mu^2}\,\phi^*\phi\ +\ \tfrac{g^2}{\mu^4}(\phi^*\phi)^2\ +\ O(g^3) \ .
\end{equation}
This action is invariant under the standard supersymmetry transformations\
$\delta_\xi=\xi^\alpha\delta_\alpha+\bar\xi_{\dot\alpha}\bar\delta^{\dot\alpha}$
with anticommuting Weyl-spinor parameters $(\xi^\alpha,\bar\xi_{\dot\alpha})$, acting as
\begin{equation} \label{susy}
\delta_\xi \phi \= \sqrt{2}\,\xi^\alpha\psi_\alpha\ ,\qquad
\delta_\xi \psi_\beta \= \im\sqrt{2} (\slashed{\pa}_{\beta\dot\alpha}\phi)\,\bar\xi^{\dot\alpha} + \sqrt{2}\,\xi_\beta F\ ,\qquad
\delta_\xi F \= \im\sqrt{2}\,\bar\xi_{\dot\alpha}\smash{\bar{\slashed{\pa}}}^{\dot\alpha\beta}\psi_\beta \ .
\end{equation}
Eliminating the auxiliary field~$F$ by its algebraic equation of motion 
\begin{equation}
F \= -\tfrac{g}{\mu^2} f_g\,\phi^* (\psi\psi)
\end{equation}
we arrive at a lagrangian
\begin{equation} \label{LagrCP1}
\begin{aligned}
\Lagr &\= -f_g^2\,\partial_m \phi\,\partial^m \phi^* 
-\tfrac{\im}{2}f_g^2 \bigl(\psi\,\slashed{\pa} \bar\psi+\bar\psi\,\bar{\slashed{\pa}} \psi\bigr)
+\im\,\tfrac{g}{\mu^2} f_g^3\,\psi\,(\phi\,\slashed{\pa}\phi^*{-}\phi^*\slashed{\pa}\phi)\,\bar\psi
-\tfrac 12 \tfrac{g}{\mu^2}f_g^4\,( \bar\psi \bar\psi )\,(\psi \psi) \\[4pt]
&\= -f_g^2\,\partial_m \phi\, \partial^m \phi^* 
-\im f_g^2\,\psi\,\slashed{\pa} \bar\psi
+2\im\,\tfrac{g}{\mu^2}f_g^3\,\psi\,\phi(\slashed{\pa}\phi^*\!)\,\bar\psi
-\tfrac 12 \tfrac{g}{\mu^2}f_g^4\,( \bar\psi \bar\psi)\,(\psi \psi)
\ +\ \textrm{total derivative}\ .
\end{aligned}
\end{equation}
The construction of the Nicolai map requires not only the $D$ term~\eqref{Kaction} in the superfield expansion of~\eqref{K}
but also its penultimate component,
\begin{equation}
\Deltc_\alpha \= \int\!\diff\theta_\alpha\,\diff^2\bar\theta \ K(\Phi^\dagger ,\Phi)
\end{equation}
and its hermitian conjugate $\ensuremath{\mathring{\bar\Delt}}^{\dot\alpha}$.
This fermionic functional computes to
\begin{equation}
\Deltc_\alpha \= -\im \sqrt{2}\,f_g^2\,(\slashed{\pa}\phi)_{\alpha\dot\alpha}\,\bar\psi^{\dot\alpha}
\ +\ \sqrt{2} f_g^2\,F^*\psi_\alpha
\ -\ \sqrt 2\,\tfrac{g}{\mu^2} f_g^3\,\phi\,\psi_\alpha(\bar\psi\bar\psi)
\end{equation}
and, after elimination of~$F$, turns into
\begin{equation}\label{MCP1}
\Delt_\alpha \= -\im \sqrt{2}\,f_g^2\,(\slashed{\pa}\phi)_{\alpha\dot\alpha}\,\bar\psi^{\dot\alpha}
\ -\  2 \sqrt 2\,\tfrac{g}{\mu^2} f_g^3\,\phi\,\psi_\alpha(\bar\psi\bar\psi)\ .
\end{equation}
It allows one to express the lagrangian as a supersymmetry variation,
\begin{equation}
\delta_\xi \Deltc_\alpha \= 2\,\xi_\alpha\,\Lagrc \qquad\textrm{or}\qquad
\delta_\xi \ensuremath{\mathring{\bar\Delt}}^{\dot\alpha} \= 2\,\bar\xi^{\dot\alpha}\,\Lagrc 
\end{equation}
up to total derivatives, which implies, most generally, 
\begin{equation}
\Lagrc \= -\tfrac14 \bigl( 
\tfrac{1+\kappa}{2}\,\delta^\alpha \Deltc_\alpha 
+ \tfrac{1-\kappa}{2}\,\bar\delta_{\dot\alpha} \ensuremath{\mathring{\bar\Delt}}^{\dot\alpha} \bigr)
\ +\ \textrm{total derivative}\ ,
\end{equation}
with a free parameter~$\kappa$ balancing the chiral against the antichiral contribution.
Clearly, the choice of $\kappa{=}1$ (or $\kappa{=}{-}1$) leads to substantial simplifications. 
Effectively, it employs only half of the supersymmetry.\footnote{
Strictly speaking, keeping $\xi^\alpha$ while putting $\bar\xi_{\dot\alpha}$ to zero
requires complexifying the Majorana superymmetry parameter and will lead to a complexified Nicolai map,
but this is inconsequential for all our purposes.}
Therefore, from now on we specialize to a ``chiral'' Nicolai map by taking $\kappa=1$.

We now have all the ingredients for setting up the Nicolai map.
Observing that $\pa_g$ commutes with the off-shell supersymmetry transformations $\delta_\xi$ in~\eqref{susy}
and with the help of
\begin{equation}
\delta_\alpha \phi = \sqrt{2}\,\psi_\alpha \und
\delta_\alpha \phi^* = 0
\end{equation}
we can engineer the coupling flow operator
\begin{equation}
R_g \=  -\tfrac{\im}{4}\sqrt{2}\int\!\diff^4{y}\!\int\!\diff^4{x}\ 
\bigl\langle\!\bigl\langle\pa_g\Deltc_\alpha(y)\ \psi^{\alpha}(x)\bigr\rangle\!\bigr\rangle\,\tfrac{\delta}{\delta\phi(x)}
\end{equation}
where the double bracket $\langle\!\langle\ldots\rangle\!\rangle$ indicates averaging over $(\psi_\alpha,\bar\psi^{\dot\alpha})$ 
as well as over $(F,F^*)$ with respect to the lagrangian~\eqref{LagrCP1c}.
Hence, this functional differential operator acts in the effective nonlocal theory for the complex boson~$(\phi,\phi^*)$.
Since $\Deltc_\alpha$ is just linear in~$(F,F^*)$, the auxiliary fields can be integrated out explicitly, leading to
\begin{equation}
\begin{aligned}
R_g &\= -\tfrac{\im}{4}\sqrt{2}\int\!\diff^4{y}\!\int\!\diff^4{x}\ 
\bigl\langle\!\bigl\langle\pa_g\Deltc_\alpha(y)\ \psi^{\alpha}(x)\bigr\rangle\!\big\rangle\,\tfrac{\delta}{\delta\phi(x)} \\[4pt]
&\= -\tfrac{1}{2} \int\!\diff^4{y}\!\int\!\diff^4{x}\ (\pa_g f_g^2)\,\slashed{\pa}_{\alpha\dot\alpha}\phi(y)\,
\bigl\langle\bar\psi^{\dot\alpha}(y)\ \psi^{\alpha}(x)\bigr\rangle\,\tfrac{\delta}{\delta\phi(x)} \\
&\qquad\! +\tfrac{\im}{2} \int\!\diff^4{y}\!\int\!\diff^4{x}\ 
\bigl(\pa_g(\tfrac{g}{\mu^2}f_g^3)+\tfrac{g}{\mu^2}f_g\,\pa_g f_g^2)\bigr)\,\phi(y)\,
\bigl\langle \psi_\alpha(y)\,\bar\psi_{\dot\alpha}(y)\,\bar\psi^{\dot\alpha}(y)\ \psi^{\alpha}(x)\bigr\rangle\,
\tfrac{\delta}{\delta\phi^(x)} \\[4pt]
&\= \tfrac{1}{\mu^2}\!\int\!\diff^4{y}\!\int\!\diff^4{x}\ f_g^3\,\Bigl\{
\phi^*\phi\,\tr\,\slashed{\pa}\phi(y)\,\bigl\langle\bar\psi(y)\ \psi(x)\bigr\rangle
-(\tfrac52 f_g{-}2)\im\,\phi(y)\,\bigl\langle (\bar\psi\,\bar\psi)(y)\,(\psi(y)\ \psi(x)) \bigr\rangle\,
\Bigr\}\,\tfrac{\delta}{\delta\phi(x)} \\[4pt]
&\ =:\ R_g^{\textrm{cl}}\ +\ R_g^{\textrm{qu}}
\end{aligned}
\end{equation}
where the single bracket $\langle\ldots\rangle$ signifies the fermionic path integral for the lagrangian~\eqref{LagrCP1}
(where $(F,F^*)$ has been removed), and the trace $\tr$ is over spinor indices.
The first term~$R_g^{\textrm{cl}}$, containing the fermion two-point function in the bosonic background but ignoring
the four-fermi interaction, has a familiar form and will produce, upon exponentiation, the tree structure of the classical Nicolai map.
The second term $R_g^{\textrm{qu}}$, featuring the fermion four-point function as well as the quartic fermion interaction
contributions to the two-point function, is of order~$\hbar$ relative to the classical term (not explicitly shown here). 
It leads to fermion-loop decorations of the classical Nicolai-map tree.
Finally, since $R_g\phi^*=0$ due to our chiral choice of~$\kappa{=}1$, we already know that
\begin{equation}
(T_g \phi^*)(x) \= \phi^*(x) \ .
\end{equation}

\section{The classical map}

\noindent
Before constructing the Nicolai map, let us test the accuracy of $R_g^{\textrm{cl}}$. 
The infinitesimal version of the free-action condition demands that $R_g^{\textrm{cl}}$ annihilates the bosonic part of
the action,
\begin{equation}
0 \ \buildrel{!}\over{=}\ \bigl(\pa_g + R_g^{\textrm{cl}}\bigr)\,S_g^{(0)} 
\= \bigl( \pa_g + R_g^{\textrm{cl}}\bigr)\,\int\!\diff^4x\ \phi\;\pa_m (f_g^2\,\pa^m\phi^*)\ .
\end{equation}
To verify this relation we need a functional relation for the classical fermionic two-point function.
The latter is just the propagator in the absence of the four-fermi interaction,
\begin{equation}
\bigl( -f_g^2\,\slashed{\pa} \ +\ 2\tfrac{g}{\mu^2} f_g^3\,\phi\,\slashed{\pa}\phi^*\bigr)(y)\,
\bigl\langle \bar\psi(y)\ \psi(x)\bigr\rangle_{\textrm{cl}} \= \delta^{(4)}(y{-}x)\ ,
\end{equation}
and therefore obeys the implicit relation
\begin{equation} \label{prop}
\bigl\langle \bar\psi(y)\ \psi(x)\bigr\rangle_{\textrm{cl}} \= -\bar{\slashed{\pa}} \Box^{-1}(y{-}x)\,f_g^{-2}(x)
\ +\ 2\tfrac{g}{\mu^2}\int\!\diff^4z\ \bigl\langle \bar\psi(y)\ \psi(z)\bigr\rangle_{\textrm{cl}}\ 
f_g^3\,\phi\,\slashed{\pa}\phi^*(z)\,\bar{\slashed{\pa}}\Box^{-1}(z{-}x)\,f_g^{-2}(x)
\end{equation}
which, upon iteration, yields the perturbative expansion of the fermion propagator.
As a consequence,
\begin{equation} \label{propadd}
\bigl\langle \bar\psi(y)\ \psi(x)\bigr\rangle_{\textrm{cl}}\ f_g^2\,\Box\phi^*(x) \=
-\delta^{(4)}(y{-}x)\;\bar{\slashed{\pa}}\phi^*(x)\ +\ 2\tfrac{g}{\mu^2}\ \bigl\langle \bar\psi(y)\ \psi(x)\bigr\rangle_{\textrm{cl}}\
f_g^3\,\phi\;\pa_m\phi^*\,\pa^m\phi^*(x)\ ,
\end{equation}
which enables us to compute
\begin{align} \label{freeactioncheck}
R_g^{\textrm{cl}}\,S_g^{(0)} &\=
\tfrac{1}{\mu^2}\!\int\!\diff^4{y}\!\int\!\diff^4{x}\ f_g^3\,\phi^*\phi\,\tr\,\slashed{\pa}\phi(y)\ 
\bigl\langle\bar\psi(y)\ \psi(x)\bigr\rangle_{\textrm{cl}}\,
\Bigl\{ \pa_m (f_g^2\,\pa^m\phi^*)\ +\ 2\tfrac{g}{\mu^2}f_g^3\,\phi^*\,\pa_m\phi\;\pa^m\phi^* \Bigr\}(x) \nn\\
&\= \tfrac{1}{\mu^2}\!\int\!\diff^4{y}\!\int\!\diff^4{x}\ f_g^3\,\phi^*\phi\,\tr\,\slashed{\pa}\phi(y)\ 
\bigl\langle\bar\psi(y)\ \psi(x)\bigr\rangle_{\textrm{cl}}\,
\Bigl\{ f_g^2\,\Box\phi^*\ -\ 2\tfrac{g}{\mu^2}f_g^3\,\phi\;\pa_m\phi^*\,\pa^m\phi^* \Bigr\}(x) \nn\\
&\= \tfrac{1}{\mu^2}\!\int\!\diff^4{y}\!\int\!\diff^4{x}\ f_g^3\,\phi^*\phi\,\tr\,\slashed{\pa}\phi(y)\
\Bigl\{ -\delta^{(4)}(y{-}x)\;\bar{\slashed{\pa}}\phi^*(x) \Bigr\} \\
&\= -\tfrac{2}{\mu^2}\int\!\diff^4{x}\ f_g^3\,\phi^*\phi\;\pa_m\phi\,\pa^m\phi^*(x)
\= \int\!\diff^4{x}\ (\pa_g f_g^2)\,\pa_m\phi\;\pa^m\phi^*(x) \= -\pa_g\,S_g^{(0)} \nn
\end{align}
and hence establish the infinitesimal free-action condition to all orders in~$g$.
The quantum contribution~$R_g^{\textrm{qu}}$ only enters (infinitesimal) determinant-matching condition 
and its higher quantum variants.

It is possible to give an explicit perturbative expansion of~$R_g^{\textrm{cl}}$.
Iterating \eqref{prop} and expanding $f_g$ one obtains, after systematic partial integrations and some combinatorics,
\begin{align} \label{Rclexp}
R_g^{\textrm{cl}} &\= 
\tfrac{1}{\mu^2}\!\int\!\diff^4{y}\!\int\!\diff^4{x}\ f_g^3\,
\phi^*\phi\,\tr\,\slashed{\pa}\phi(y)\,\bigl\langle\bar\psi(y)\ \psi(x)\bigr\rangle_{\textrm{cl}}\,\tfrac{\delta}{\delta\phi(x)} \nn\\
&\= \tfrac1{\mu^2}\!\int\!\cdots\!\int \tfrac12\tr\,\phi\,\bigl\{
Z_1 -2\tfrac{g}{\mu^2}Z_2 +3\tfrac{g^2}{\mu^4}Z_3 -4\tfrac{g^3}{\mu^6}Z_4 \pm\ldots\bigr\}\,\bigl\{
1 -2\tfrac{g}{\mu^2}Z_1 +3\tfrac{g^2}{\mu^4}Z_2 -4\tfrac{g^3}{\mu^6}Z_3 \pm\ldots \bigr\}^{-1} \tfrac{\delta}{\delta\phi} \nn\\
&\= \sum_{\bf n} \frac{g^{n-1}}{\mu^{2n}}\,(-1)^{n+s} n_s\prod_{i=1}^{s-1}(n_i{+}1) \int\!\cdots\!\int \tfrac12\tr\,
\phi\;Z_{n_s}\,Z_{n_{s-1}}\,\ldots Z_{n_2}\,Z_{n_1}\,\tfrac{\delta}{\delta\phi} \\
&\= \!\int\!\cdots\!\int \tfrac12\tr\,\phi\;\bigl\{
\tfrac{1}{\mu^2}Z_1\ -\ \tfrac{g}{\mu^4}( 2Z_2-2Z_1^2 )\ +\ \tfrac{g^2}{\mu^6}( 3Z_3-3Z_1Z_2-4Z_2Z_1+4Z_1^3 )\ \mp\ldots
\bigr\}\,\tfrac{\delta}{\delta\phi} \nn\\
&\ =:\ r_1\ +\ g\,r_2\ +\ g^2 r_3\ +\ \ldots \nn
\end{align}
with the abbreviation
\begin{equation}
Z_i \ :=\ \phi^i\,\slashed{\pa}\bigl( {\phi^*}^i\bar{\slashed{\pa}}\Box^{-1}\bigr)
\= \phi^i\,\slashed{\pa}{\phi^*}^i\,\bar{\slashed{\pa}}\Box^{-1} \ +\ (\phi^*\phi)^i
\end{equation}
and spacetime arguments having been suppressed. The sum runs over multi-indices
\begin{equation}
{\bf n} \= (n_1,n_2,\ldots,n_s)\ ,\qquad n_i \in\NN\ ,\qquad {\textstyle\sum}_i n_i=n\ ,\qquad 1\le s\le n\ ,
\end{equation}
so that each term features $s{+}1$ integrations,
and $r_n$ is a sum over multi-indices of fixed absolute value~$n$
and lengths~$s$ between 1 and~$n$.

We employ the universal formula~\cite{LR1} to compute the Nicolai map perturbatively as
a $g$-ordered exponential of ${-}\!\int\! R_g$ acting on~$\phi$.
We first compute the classical part $T_g^{(0)}\phi$ based on $R_g^{\textrm{cl}}$.
The universal formula allows for a pairwise combination of terms proportional to 
\begin{equation}
\ell\,r_k\,\phi \ -\ r_{k-\ell}\,r_\ell\,\phi \qquad\textrm{for}\quad k>\ell=1,2,3,\ldots \ .
\end{equation}
Picking from the multi-index sums in~\eqref{Rclexp} the contributions
\begin{equation}
r_k: (\ell,k_2,k_3,\ldots,k_s)\ ,\qquad
r_{k-\ell}: (k_2,k_3,\ldots,k_s)\ ,\qquad
r_\ell: (\ell) 
\qquad\textrm{with}\quad {\textstyle\sum}_i k_i=k{-}\ell 
\end{equation}
the coefficients in~\eqref{Rclexp} conspire with the rule $\frac{\pa}{\pa\phi}(\phi\,Z_\ell)=(\ell{+}1)Z_\ell$ to
a partial cancellation,
\begin{equation} \label{partcancel}
\bigl(\ell\,r_k\ -\ r_{k-\ell}\,r_\ell\bigr)\,\phi \=
\tfrac{(-1)^{k+s}}{\mu^{2k}}\,\ell(\ell{+}1) k_s\prod_{i=2}^{s-1}(k_i{+}1) \!\int\!\cdots\!\int \Bigl\{
\tfrac12\tr\bigl( Z_{k_s}\ldots Z_{k_2} Z_\ell \bigr)\ -\ 
\tfrac12\tr\bigl( Z_{k_s}\ldots Z_{k_2} \bigr)\,\tfrac12\tr\bigl( Z_\ell) \Bigr\}
\end{equation}
with $\tfrac12\tr Z_\ell=\phi^\ell\pa_m({\phi^*}^\ell\pa^m\Box^{-1})$.
The obstruction to a complete cancellation is proportional to
\begin{equation} \label{obstruction}
\phi^\ell\,\Sigma_\ell\ :=\ Z_\ell-\tfrac12\tr Z_\ell \= 
\phi^\ell\,\slashed{\pa}\bigl( {\phi^*}^\ell\bar{\slashed{\pa}}\Box^{-1}\bigr)
\ -\ \phi^\ell\,\pa_m\bigl({\phi^*}^\ell\pa^m\Box^{-1}\bigr)
\= \phi^\ell\,\sigma^{mn}\,\pa_m{\phi^*}^\ell\pa_n\Box^{-1}
\end{equation}
at the root of each tree.
It is not hard to see that every multi-index contribution to~$T_g^{(0)}\phi$ is part of precisely one such pairing,
except for those of length $s{=}1$, which arise from a single action of~$R_g^{\textrm{cl}}$ on~$\phi$. 
These unpaired contributions can be summed to
\begin{align} \label{Tone}
\bigl(T_g^{(0)}\phi\bigr)(x)\big|_{s=1} &\=
\phi(x)\ -\int_0^g\!\diff g'\ R_{g'}^{\textrm{cl}}\,\phi(x)\big|_{s=1} \nn\\
&\= \phi(x)\ -\int\!\diff^4y\ \tfrac12\tr\,\phi(y)\,\bigl\{ 
\tfrac{g}{\mu^2}Z_1 -\tfrac{g^2}{\mu^4}Z_2 +\tfrac{g^3}{\mu^6}Z_3\mp\ldots\bigr\}(y,x) \nn\\
&\= \phi(x)\ + \int\!\diff^4y\ \bigl\{
2\tfrac{g}{\mu^2}\phi^*\phi-3\tfrac{g^2}{\mu^4}(\phi^*\phi)^2+4\tfrac{g^3}{\mu^6}(\phi^*\phi)^3\mp\ldots\bigr\}(y)\,
\tfrac12\tr\,\slashed{\pa}\phi(y)\,\bar{\slashed{\pa}}\Box^{-1}(y{-}x) \nn\\
&\=\phi(x)\ -\int\!\diff^4y\ (f_g^2-1)\,\tfrac12\tr\,\slashed{\pa}\phi(y)\,\bar{\slashed{\pa}}\Box^{-1}(y{-}x) \\
&\=-\int\!\diff^4y\ f_g^2\;\pa_m\phi(y)\,\pa^m\Box^{-1}(y{-}x)\ , \nn
\end{align}
where we made use of \ $\phi+\int\pa_m\phi\,\pa^m\Box^{-1}=0$.
The (finite) free-action condition is clearly satisfied,
\begin{equation}
S_0^{(0)}[T_g^{(0)}\phi] \= \!\int\!\diff^4x\;\bigl(T_g^{(0)}\phi\bigr)(x)\,\Box\phi^*(x)
\= \!\int\!\diff^4x\;\bigl(T_g^{(0)}\phi\bigr)(x)\big|_{\textrm{length=1}}\,\Box\phi^*(x)
\= -\!\int\!\diff^4x\;f_g^2\,\pa_m\phi\,\pa^m\phi^*(x)
\end{equation}
since the obstruction~\eqref{obstruction} kills $\Box\phi^*$.
In fact, one could have guessed the form of~\eqref{Tone} directly from the free-action condition.
However, this condition fixes merely the length-one part of the classical Nicolai map.
Only in the absence of the obstruction~\eqref{obstruction}, when the spinor tace trivializes,
the classical Nicolai map collapses to the almost-local part with single-edged trees.

For the theory at hand, however, the multi-index contributions with length $s{>}1$
correspond to trees with more than one edge and do not vanish. 
Their evaluation is explicit, combining the Stirling-number coefficients in the universal formula
(expressing $T_g\phi$ in terms of $r_k$) with the multiindex coefficients in~\eqref{Rclexp}
(expressing $r_k$ in terms of $Z_\ell$).  
Let us support these claims by a direct computation of the classical map to third order.
To streamline notation, we abbreviate
\begin{equation}
[ X ] := \tfrac12\tr\,X \ ,
\end{equation}
absorb $\mu$ into~$g$ and suppress all spacetime arguments. 
Then, we proceed:
\begin{align}
T_g^{(0)}\phi &\=
\phi\ -\ g\,r_1\,\phi\ -\ \tfrac12 g^2 \bigl( r_2-r_1^2 \bigr)\,\phi\ -\ 
\tfrac16 g^3 \bigl( 2\,r_3-2\,r_2r_1-r_1r_2+r_1^3 \bigr)\,\phi\ +\ O(g^4) \nn\\[4pt]
&\= \phi\ -\ g\smallint[\phi\,Z_1]\ -\ 
\tfrac12 g^2 \bigl(-2\smallint[\phi Z_2]
+2\smallint\!\!\smallint[\phi Z_1^2] - 2\smallint\!\!\smallint[\phi Z_1][ Z_1] \bigr) \\
&\ \ - \tfrac16 g^3 \bigl( 
6\smallint[\phi Z_3]-6\smallint\!\!\smallint[\phi Z_1Z_2]
-8\smallint\!\!\smallint[\phi Z_2Z_1]+8\smallint\!\!\smallint\!\!\smallint[\phi Z_1^3] \nn\\
&\qquad\quad\!+ 8\smallint\!\!\smallint[\phi Z_2][ Z_1]
-8\smallint\!\!\smallint\!\!\smallint[\phi Z_1^2][ Z_1]
+6\smallint\!\!\smallint[\phi Z_1][ Z_2]
-4\smallint\!\!\smallint\!\!\smallint[\phi Z_1][ Z_1^2]
+4\smallint\!\!\smallint\!\!\smallint[\phi Z_1][ Z_1][ Z_1] \bigr) \nn\\
&\qquad\quad\!- 2\smallint\,[ \begin{smallmatrix}
\smallint[\phi Z_1] \\ \smallint \phi Z_1 \end{smallmatrix}\,\phi^{-1}Z_1]
+2\smallint\! \begin{smallmatrix}
\smallint[\phi Z_1] \\ \smallint [\phi Z_1]\end{smallmatrix}\,[\phi^{-1}Z_1]
\ +\ O(g^4)\ . \nn
\end{align}
In the two terms of the last line, the (hidden) spacetime arguments are not arranged linearly,
but show a forked structure: $r_1$ also acts on the right~$Z_1$ in $[\phi Z_1Z_1]$ and $[\phi Z_1][ Z_1]$
via $\tfrac{\pa}{\pa\phi}Z_\ell=\ell\phi^{-1}Z_\ell$ to produce a branched tree diagram.
It is easy to identify
\begin{equation}
T_g^{(0)}\phi\big|_{s=1} \=
\phi\ -\ g\smallint[\phi\,Z_1]\ +\ g^2\!\smallint[\phi Z_2]\ -\ g^3\!\smallint[\phi Z_3]\ +\ O(g^4)
\end{equation}
and to observe the pairwise partial cancellations in
\begin{equation}
\begin{aligned}
T_g^{(0)}\phi\big|_{s>1} &\=
-g^2 \smallint\!\!\smallint [\phi Z_1\,\phi\,\Sigma_1]
+g^3 \smallint\!\!\smallint [\phi Z_1\,\phi^2 \Sigma_2]
+\tfrac43g^3 \smallint\!\!\smallint [\phi Z_2\,\phi\,\Sigma_1]  
-\tfrac43g^3 \smallint\!\!\smallint\!\!\smallint[\phi Z_1^2\phi\,\Sigma_1] \\ &\quad\ \ 
+\tfrac23g^3 \smallint\!\!\smallint\!\!\smallint [\phi Z_1][ Z_1\,\phi\,\Sigma_1] 
+\tfrac13g^3 \smallint\,[ \begin{smallmatrix} \smallint[\phi Z_1] \\ \smallint \phi Z_1 \end{smallmatrix}\ \Sigma_1]
\ +\ O(g^4)\ . 
\end{aligned}
\end{equation}
With the help of partial integrations and $\Sigma_\ell=\ell{\phi^*}^{\ell-1}\Sigma_1$, 
one may cast the expansion into a different form,
\begin{equation} \label{Texplicit}
\begin{aligned}
T_g^{(0)}\phi &\=
-\smallint f_g^2\,\pa\phi\cdot\pa\Box^{-1}
-2g \sum_{k=1}^\infty\sum_{\ell=0}^\infty (-1)^{k+\ell}\tfrac{k(k+1)}{k+\ell+1} \smallint\!\!\smallint
[(g\,\phi^*\phi)^k\slashed{\pa}\phi\,\bar{\slashed{\pa}}\Box^{-1}(g\,\phi^*\phi)^\ell\phi\;\Sigma_1] \\
&\ +\ \tfrac43g^3 \smallint\!\!\smallint\!\!\smallint 
[\phi^*\phi\,(2\,\slashed{\pa}\phi\,\bar{\slashed{\pa}}-\pa\phi\cdot\pa)\Box^{-1}
\phi\,\slashed{\pa}\phi^*\bar{\slashed{\pa}}\Box^{-1}\phi\;\Sigma_1]
\ +\ \tfrac43g^3 \smallint\,[ \begin{smallmatrix} 
\smallint \phi^*\phi\,\pa\phi\cdot\pa\Box^{-1} \\
\smallint \phi^*\phi\,\slashed{\pa}\phi\,\bar{\slashed{\pa}}\Box^{-1}
\end{smallmatrix}\,\Sigma_1] \ +\ O(g^4)\ ,
\end{aligned}
\end{equation}
where the dots signify Lorentz contractions of vector indices, and we remind the reader of
\begin{equation}
\Sigma_1 \= \sigma^{mn}\,\pa_m\phi^*\pa_n\Box^{-1}\ .
\end{equation}
The first term in~\eqref{Texplicit} is the complete length-one part of the map, 
and in the second term (which starts at order~$g^2$) we have summed all length-two contributions. 
This perturbative tree expansion may be pushed systematically to any desired order.
Performing the spinorial traces will Lorentz-contract vector indices on partial derivatives in all possible ways
along paths inside each tree diagram (except for the two indices next to the root). 
These `long-distance' index correlations prohibit the collapse of the classical Nicolai map 
to the almost-local expression~\eqref{Tone}.

\section{The quantum map}

\noindent
Having achieved a thorough understanding of the classical Nicolai map, 
we now turn to the quantum part,
\begin{equation}
T_g^{\textrm{qu}}\phi\ :=\ 
T_g\,\phi-T_g^{(0)}\phi \=  
\hbar\,T_g^{(1)}\phi + \hbar^2 T_g^{(2)}\phi + \hbar^3 T_g^{(3)}\phi + \ldots\ ,
\end{equation}
but will restrict ourselves here to order~$g^2$ and suppress~$\hbar$ again. With
\begin{align} \label{Rqu}
R_g^{\textrm{qu}} &\=
\tfrac{1}{\mu^2}\!\int\!\diff^4{y}\!\int\!\diff^4{x}\ f_g^3\,\Bigl\{
\phi^*\phi\,\tr\,\slashed{\pa}\phi(y)\,\bigl\langle\bar\psi(y)\ \psi(x)\bigr\rangle_{\textrm{qu}}
-(\tfrac52f_g{-}2)\im\,\phi(y)\,\bigl\langle (\bar\psi\,\bar\psi)(y)\,(\psi(y)\ \psi(x)) \bigr\rangle\,
\Bigr\}\,\tfrac{\delta}{\delta\phi(x)} \nn \\
&\ =:\ r'_1 + g\,r'_2 + g^2 r'_3 + \ldots
\end{align}
we have
\begin{equation}
T_g^{\textrm{qu}}\phi \= -g\,r'_1\phi - \tfrac12g^2\bigl( r'_2-r_1r'_1-r'_1r_1)\,\phi + O(g^3)\ .
\end{equation}
In each $r'_k$, the fermionic 4-point function (with three points identified) 
and also the quantum part of the two-point function can be expanded in the number of fermion loops 
(and thus powers of~$\hbar$).
The ensuing graphical representation of the full Nicolai map to order~$g^2$ looks as follows,
\newcommand\scale{.70}
\begin{align} \label{graphs}
T_g\phi\=
\scalebox{\scale}{
\begin{tikzpicture}[baseline={([yshift=-1ex]current bounding box.center)},thick]
\node at (0,0)[circle,fill,inner sep = 2pt] {};
\draw[gluon](1,0)--(0,0);
\end{tikzpicture} 
}
 & \ +\ 
g \
\scalebox{\scale}{
\begin{tikzpicture}[baseline={([yshift=-1ex]current bounding box.center)},thick]
\node at (0,0)[circle,fill,inner sep = 2pt] {};
\draw (1,0)--(0,0);
\draw[gluon] (1+0.75, 0.75)--(1,0);
\draw[gluon] (1+1,0)--(1,0);
\draw[gluon] (1+0.75,-0.75)--(1,0);
\end{tikzpicture} 
}
 \ +\ 
g ^2\
\left(
	\scalebox{\scale}{
	\begin{tikzpicture}[baseline={([yshift=-1ex]current bounding box.center)},thick]
	\node at (-0.25,0)[circle,fill,inner sep = 2pt] {};
	\draw (2,0)--(-0.25,0);
	\draw[gluon] (2+0.75, 0.75)--(2,0);
	\draw[gluon] (2+1,0)--(2,0);
	\draw[gluon] (2+0.75,-0.75)--(2,0);
	\draw[gluon] (1+0.5,-0.9)--(1,0);
	\draw[gluon] (1-0.5,-0.9)--(1,0);
	\phantom{ \draw[gluon] (1-0.5,0.9)--(1,0); }
	\end{tikzpicture} 
	}
	\ +\ 
	\scalebox{\scale}{
	\begin{tikzpicture}[baseline={([yshift=-1ex]current bounding box.center)},thick]
	\node at (0,0)[circle,fill,inner sep = 2pt] {};
	\draw (1,0)--(0,0);
	\draw[gluon] (1+0.75, 0.75)--(1,0);
	\draw[gluon] (1+0.90, 0.40)--(1,0);
	\draw[gluon] (1+1,0)--(1,0);
	\draw[gluon] (1+0.90, -0.40)--(1,0);
	\draw[gluon] (1+0.75,-0.75)--(1,0);
	\end{tikzpicture} 
	}
\right)
\ + \ \cdots
\nn \\
& \ +\ 
\hbar\,g \
\scalebox{\scale}{
\begin{tikzpicture}[baseline={([yshift=-1ex]current bounding box.center)},thick]
\node at (0,0)[circle,fill,inner sep = 2pt] {};
\draw (0.9,0)--(0,0); 
\draw (0.9,0)--(0.9-  0.75*0.288 , 0.75*1/2); 
\draw (0.9,0)--(0.9+ 0.75*0.288 ,0.75*1/2); 
\draw[gluon] (0.9+0.9,0)--(0.9,0); 
\draw [domain=-30:180+30] plot ( { 0.75* cos(\x)/3 + 0.9  } , {   0.75* sin(\x)/3+0.75* 2/3 } );
\phantom{
\draw [domain=-30:180+30] plot ( { 0.75* cos(\x)/3 + 0.9  } , {-   0.75* sin(\x)/3-0.75* 2/3 } );
}
\end{tikzpicture} 
}
  \ +\ 
\hbar\,g^2 
\left(
	\scalebox{\scale}{
	\begin{tikzpicture}[baseline={([yshift=-1ex]current bounding box.center)},thick]
	\node at (-1,0)[circle,fill,inner sep = 2pt] {};
	\draw (0.7,0)--(-1,0); 
	\draw (0.7,0)--(0.7-  0.75*0.288 , 0.75*1/2); 
	\draw (0.7,0)--(0.7+ 0.75*0.288 ,0.75*1/2); 
	\draw[gluon] (0.7+0.9,0)--(0.7,0); 
	\draw [domain=-30:180+30] plot ( { 0.75* cos(\x)/3 + 0.7  } , {   0.75* sin(\x)/3+0.75* 2/3 } );
	\phantom{
	\draw [domain=-30:180+30] plot ( { 0.75* cos(\x)/3 + 0.7  } , {-   0.75* sin(\x)/3-0.75* 2/3 } );
	}
	\draw[gluon] (0.5,-0.9)--(0,0);
	\draw[gluon] (-0.5,-0.9)--(0,0);
	\phantom{ \draw[gluon] (1-0.5,0.9)--(1,0); }
	\end{tikzpicture} 
	} 
	 +
	\scalebox{\scale}{
	\begin{tikzpicture}[baseline={([yshift=-1ex]current bounding box.center)},thick]
	\node at (0,0)[circle,fill,inner sep = 2pt] {};
	\draw (0.7,0)--(0,0); 
	\draw (0.7,0)--(0.7-  0.65*0.288 , 0.65*1/2); 
	\draw (0.7,0)--(0.7+ 0.65*0.288 ,0.65*1/2); 
	\draw[gluon] (0.7+0.9,0)--(0.7,0); 
		\draw [domain=-30:180+30] plot ( { 0.65* cos(\x)/3+0.7  } , {   0.65* sin(\x)/3+0.65* 2/3 } );
		\draw[gluon] (0.7+0.4  ,1.1 ) --(0.7,0.65 );
		\draw[gluon] (0.7-0.4  ,1.1  ) --(0.7,0.65 );
	\phantom{
		\draw[gluon] (0.7+0.4  ,-1.1 ) --(0.7,-0.65 );
	}
	\end{tikzpicture} 
	}
		+
	\scalebox{\scale}{
	\begin{tikzpicture}[baseline={([yshift=-1ex]current bounding box.center)},thick]
	\node at (0,0)[circle,fill,inner sep = 2pt] {};
	\draw (1.2,0)--(0,0);
	\draw[gluon] (2.15-0.2, 0.75)--(1.4-0.2,0);
	\draw[gluon] (2.4-0.2,0)--(1.4-0.2,0);
	\draw[gluon] (2.15-0.2,-0.75)--(1.4-0.2,0);
		\draw (0.7,0)--(0.7-  0.75*0.288 , 0.75*1/2); 
		\draw (0.7,0)--(0.7+ 0.75*0.288 ,0.75*1/2); 
		\draw [domain=-30:180+30] plot ( { 0.75* cos(\x)/3+0.7  } , {   0.75* sin(\x)/3+0.75* 2/3 } );
	\end{tikzpicture} 
	}	 
	+ 
	\scalebox{\scale}{
		\begin{tikzpicture}[baseline={([yshift=-1ex]current bounding box.center)},thick]
		\node at (-1,0)[circle,fill,inner sep = 2pt] {};
		\draw (0,0)--(-1,0); 
		\draw (0,0)--( -  0.75*0.288 , 0.75*1/2); 
		\draw (0,0)--( + 0.75*0.288 ,0.75*1/2); 
		\draw[gluon] (0.9,0)--(0,0); 
		\draw [domain=-30:180+30] plot ( { 0.75* cos(\x)/3  } , {   0.75* sin(\x)/3+0.75* 2/3 } );
		\phantom{
		\draw [domain=-30:180+30] plot ( { 0.75* cos(\x)/3  } , {-   0.75* sin(\x)/3-0.75* 2/3 } );
		}
		\draw[gluon] (0.5,-0.9)--(0,0);
		\draw[gluon] (-0.5,-0.9)--(0,0);
		\phantom{ \draw[gluon] (1-0.5,0.9)--(1,0); }
		\end{tikzpicture} 
		} 
\right)
\ +\ \cdots
\nn \\
&  \
\phantom{ \ +\ 
			\hbar g \
			\scalebox{\scale}{
			\begin{tikzpicture}[baseline={([yshift=-1ex]current bounding box.center)},thick]
			\node at (0,0)[circle,fill,inner sep = 2pt] {};
			\draw (1,0)--(0,0); 
			\draw (1,0)--(1-  0.75*0.288 , 0.75*1/2); 
			\draw (1,0)--(1+ 0.75*0.288 ,0.75*1/2); 
			\draw[gluon] (1+0.9,0)--(1,0); 
			\draw [domain=-30:180+30] plot ( { 0.75* cos(\x)/3+1  } , {   0.75* sin(\x)/3+0.75* 2/3 } );
			\phantom{
			\draw [domain=-30:180+30] plot ( { 0.75* cos(\x)/3+1  } , {-   0.75* sin(\x)/3-0.75* 2/3 } );
			}
			\end{tikzpicture} 
			} 
	}
\ +\ 
\hbar^2 g^2 
\left(
	\scalebox{\scale}{
	\begin{tikzpicture}[baseline={([yshift=-1ex]current bounding box.center)},thick]
	\node at (-1,0)[circle,fill,inner sep = 2pt] {};
	\draw (1,0)--(-1,0); 
	\draw (1,0)--(1-  0.75*0.288 , 0.75*1/2); 
	\draw (1,0)--(1+ 0.75*0.288 ,0.75*1/2); 
	\draw[gluon] (1+0.9,0)--(1,0); 
	\draw [domain=-30:180+30] plot ( { 0.75* cos(\x)/3+1  } , {   0.75* sin(\x)/3+0.75* 2/3 } );
	\draw [domain=-30:180+30] plot ( { 0.75* cos(\x)/3   } , {   0.75* sin(\x)/3+0.75* 2/3 } );
	\draw (0,0)--( -  0.75*0.288 , 0.75*1/2); 
	\draw (0,0)--( 0.75*0.288 ,0.75*1/2); 
	\phantom{
	\draw [domain=-30:180+30] plot ( { 0.75* cos(\x)/3+1  } , {-   0.75* sin(\x)/3-0.75* 2/3 } );
	} 
	\end{tikzpicture} 
	} 
	\ +\ 
	\scalebox{\scale}{
	\begin{tikzpicture}[baseline={([yshift=-1ex]current bounding box.center)},thick]
	\node at (0,0)[circle,fill,inner sep = 2pt] {};
	\draw (1,0)--(0,0); 
	\draw (1,0)--(1-  0.65*0.288 , 0.65*1/2); 
	\draw (1,0)--(1+ 0.65*0.288 ,0.65*1/2); 
	\draw[gluon] (1+0.9,0)--(1,0); 
		\draw [domain=-30:180+30] plot ( { 0.65* cos(\x)/3+1  } , {   0.65* sin(\x)/3+0.65* 2/3 } );
		\draw (1,0.65)--(1-  0.65*0.288 , 0.65*1/2 + 0.65 ); 
		\draw (1,0.65)--(1+ 0.65*0.288 , 0.65*1/2 + 0.65 );  
		\draw [domain=-30:180+30] plot ( { 0.65* cos(\x)/3+1  } , {  0.65+ 0.65* sin(\x)/3+0.65* 2/3 } );
	\phantom{
		\draw [domain=-30:180+30] plot ( { 0.65* cos(\x)/3+1  } , {  - 0.65- 0.65* sin(\x)/3-0.65* 2/3 } );
	}
	\end{tikzpicture} 
	}
	\ +\ 
	\scalebox{\scale}{
	\begin{tikzpicture}[baseline={([yshift=-1ex]current bounding box.center)},thick]
	\node at (-1/3,0)[circle,fill,inner sep = 2pt] {};
	\draw (1,0)--(-1/3,0);  
	\draw[gluon] (1+0.9,0)--(1,0); 
		\draw [domain=0:360] plot ( {   (  2 +  cos(\x)  )  /3} , {    sin(\x)  /3  } );
	\end{tikzpicture} 
	}
\right)
\ + \ \cdots
\nn \\
& \qquad\qquad\qquad\qquad\qquad\qquad\qquad\qquad \ +\ O(\hbar^3 g^3) \ .
\end{align}
Here, the thick dot at the left end of each diagram stands for the argument $x$ of the map,
other vertex positions are integrated over.
Solid lines are free fermion propagators $\bar{\slashed\pa}\Box^{-1}$,
and wavy lines represent bosonic field insertions $\phi$ or $\phi^*$. 
One of the bosonic legs emanating from each vertex not sourcing a loop carries a derivative (not shown). 
For the full `Nicolai rules', one of course needs to add spinor traces and weight factors.
All diagrams shown above already appear in the first application of $R_g$ on~$\phi$.
We see that in the $\hbar$ expansion of the map an $r$-loop contribution arises first at order $g^r$, 
so that at each given order in perturbation theory only a finite number of diagrams contribute.

To the orders $g$ and $g^2$ of~$T_g^{\textrm{qu}}\phi$, only one- and two-loop diagrams contribute.
All tadpole diagrams are proportional to $\pa_m\Box^{-1}(0)$ and thus put to zero in dimensional 
regularization.\footnote{
Their leading unregularized contribution has been computed in~\cite{CLR}.}
Therefore, only the third graphs in the second and third lines of~\eqref{graphs} survive 
to the desired order. We have to evaluate
\begin{align}
&\bigl\langle (\bar\psi\,\bar\psi)(y)\,(\psi(y)\ \psi(x)) \bigr\rangle \=
-2\im\,\bigl\langle \bar\psi^{\dot\alpha}(y)\,\psi^\alpha(y) \bigr\rangle\
\bigl\langle \bar\psi_{\dot\alpha}(y)\,\psi_\alpha(x) \bigr\rangle \nn\\
&\qquad\qquad\qquad\qquad\qquad\quad
-\ 2\im\,\tfrac{g}{\mu^2}\!\int\!\diff^4z\,
\bigl\langle \bar\psi_{\dot\alpha}(y)\,\psi_\beta(z) \bigr\rangle\,
\bigl\langle \bar\psi^{\dot\alpha}(y)\,\psi^\beta(z) \bigr\rangle\,
\bigl\langle \psi^\alpha(y)\,\bar\psi^{\dot\beta}(z) \bigr\rangle\,
\bigl\langle \bar\psi_{\dot\beta}(z)\,\psi_\alpha(x) \bigr\rangle\
+\ O(g^2) \nn \\
&=\ -2\im\,\tr\,\Bigl\{ \bar{\slashed{\pa}} \Box^{-1}(0)\,f_g^{-2}(y)
\, +\, 2\tfrac{g}{\mu^2}\!\int\!\!\diff^4z\ \bar{\slashed{\pa}} \Box^{-1}(y{-}z)\,
\phi\,\slashed{\pa}\phi^*(z)\,\bar{\slashed{\pa}}\Box^{-1}(z{-}y)\,+\, O(g^2) \Bigr\}\,
\Bigl\{\slashed{\pa}\Box^{-1}(y{-}x)\, +\, O(g) \Bigr\} \nn\\
&\quad\; -8\im\,\tfrac{g}{\mu^2}\!\int\!\diff^4z\ 
\pa\Box^{-1}(y{-}z)\cdot\pa\Box^{-1}(y{-}z)\ \pa\Box^{-1}(y{-}z)\cdot\pa\Box^{-1}(z{-}x)
\ +\ O(g^2)\ .
\end{align}
Regularizing $\bar{\slashed{\pa}}\Box^{-1}(0)\to 0$ we arrive at
$T_g^{\textrm{qu}}\phi=-\tfrac12 g^2 r'_2\,\phi+O(g^3)$, thus
\begin{align} \label{Tqu}
\bigl(T_g^{(1)}\phi\bigr)(x) &\= 4\tfrac{g^2}{\mu^4}\!\int\!\diff^4y\!\int\!\diff^4z\ 
[\phi(y)\,\bar{\slashed{\pa}} \Box^{-1}(y{-}z)\,
\phi\,\slashed{\pa}\phi^*(z)\,\bar{\slashed{\pa}}\Box^{-1}(z{-}y)\,\slashed{\pa}\Box^{-1}(y{-}x)] 
\ +\ O(g^3)\ ,\\
\bigl(T_g^{(2)}\phi\bigr)(x) &\= 4\tfrac{g^2}{\mu^4}\!\int\!\diff^4y\!\int\!\diff^4z\
\phi(y)\ \pa\Box^{-1}(y{-}z)\cdot\pa\Box^{-1}(y{-}z)\ \; \pa\Box^{-1}(y{-}z)\cdot\pa\Box^{-1}(z{-}x)
\ +\ O(g^3)\ .
\end{align}

To verify the first expression we investigate the one-loop condition of the Nicolai map,
{\it i.e.\/} the (generalized) determinant-matching condition~\eqref{matching01},
\begin{equation} \label{detmatch}
S_0^{(0)}\bigl[ T_g^{(1)}\phi\bigr]\ -\ \im\,\Tr\,\ln\,\frac{\delta T_g^{(0)}\!\phi}{\delta\phi}
\ \buildrel{!}\over{=}\ S_g^{(1)}[\phi]\  .
\end{equation}
The first term can be written in this way because the free action is linear in~$\phi$,
and $\phi^*$ is not transformed. With the input of~\eqref{Tqu}, we have
\begin{equation}
\begin{aligned}
S_0^{(0)}\bigl[ T_g^{(1)}\phi\bigr] &\= 
-\int\!\diff^4x\ f_g^2\,\pa_m\phi^*\,\pa^m\bigl(T_g^{(1)}\phi\bigr)(x) 
\= \int\!\diff^4x\ \phi^*\,\Box \bigl(T_g^{(1)}\phi\bigr)(x) \ +\ O(g^3) \\
&\= 4\tfrac{g^2}{\mu^4}\!\int\!\diff^4y\!\int\!\diff^4z\ 
[\phi(y)\,\bar{\slashed{\pa}} \Box^{-1}(y{-}z)\,\phi\,\slashed{\pa}\phi^*(z)\,
\bar{\slashed{\pa}}\Box^{-1}(z{-}y)\,\slashed{\pa}\phi^*(y)] \ +\ O(g^3) \ .
\end{aligned}
\end{equation}
The second term in~\eqref{detmatch} comes from the functional Jacobian
of the classical map~\eqref{Texplicit},
\begin{align}
\frac{\delta \bigl(T_g^{(0)}\!\phi\bigr)(x)}{\delta\phi(y)} &\=
f_g^2(y)\,\delta^{(4)}(y{-}x)\ -\
 2\tfrac{g}{\mu^2}\,[\phi\,\slashed{\pa}\phi^*(y)\bar{\slashed{\pa}}\Box^{-1}(y{-}x)]\ +\ 
3\tfrac{g^2}{\mu^4}\,[\phi^2 \slashed{\pa}{\phi^*}^2(y)\bar{\slashed{\pa}}\Box^{-1}(y{-}x)] \nn \\
&\ -\ \tfrac{g^2}{\mu^4}\!\int\!\diff^4z\, \bigl\{
2\,[\phi\,\slashed{\pa}\phi^*(y)\,\bar{\slashed{\pa}}\Box^{-1}(y{-}z)\,\phi(z)\,\Sigma_1(z,x)]
+[\phi^2\slashed{\pa}\phi^*(z)\,\bar{\slashed{\pa}}\Box^{-1}(z{-}y)\,\Sigma_1(y,x)] \bigr\}
\ +\ O(g^3) \nn \\
&\ =:\ \delta^{(4)}(y{-}x)\ +\ \tfrac{g}{\mu^2}\,J_1(y,x)\ +\ \tfrac{g^2}{\mu^4}\,J_2(y,x)\ +\ O(g^3)\ . 
\end{align}
With this and $\Sigma_1(x,x)=0$ we find
\begin{align}
-\ \im\,\Tr\,\ln\,\frac{\delta T_g^{(0)}\!\phi}{\delta\phi} &\=
\int\!\diff^4x\ \Bigl\{ \tfrac{g}{\mu^2}\,J_1(x,x) + \tfrac{g^2}{\mu^4}\,J_2(x,x) \Bigr\}
-\tfrac12\int\!\diff^4x\!\int\!\diff^4y\ \tfrac{g^2}{\mu^4}\,J_1(x,y)\,J_1(y,x)\ +\ O(g^3) \nn\\
&\= {}-2\tfrac{g^2}{\mu^4}\!\int\!\diff^4x\!\int\!\diff^4y\ 
[\phi\,\slashed{\pa}\phi^*(x)\,\bar{\slashed{\pa}}\Box^{-1}(x{-}y)\,
\phi\,\sigma^{mn}\pa_m\phi^*(y)\,\pa_n\Box^{-1}(y{-}x)] \\ 
&\qquad -2\tfrac{g^2}{\mu^4}\!\int\!\diff^4x\!\int\!\diff^4y\ 
[\phi\,\slashed{\pa}\phi^*(x)\,\bar{\slashed{\pa}}\Box^{-1}(x{-}y)]\,
[\phi\,\slashed{\pa}\phi^*(y)\,\bar{\slashed{\pa}}\Box^{-1}(y{-}x)]\ +\ O(g^3) \nn\\
&\= {}-2\tfrac{g^2}{\mu^4}\!\int\!\diff^4x\!\int\!\diff^4y\ 
[\phi\,\slashed{\pa}\phi^*(x)\,\bar{\slashed{\pa}}\Box^{-1}(x{-}y)\,
\phi\,\slashed{\pa}\phi^*(y)\,\bar{\slashed{\pa}}\Box^{-1}(y{-}x)]\ +\ O(g^3) \nn
\end{align}
and observe that a length-one and a length-two contribution of~$T_g^{(0)}\phi$
combine to a simple result.
The conjugate field~$\phi^*$ does not contribute to the Jacobian because
of $T_g\phi^*=\phi^*$.
Finally, the fermion determinant computes to~\footnote{
The factor of $\tfrac12$ is due to the Majorana nature of the fermions. }
\begin{equation}
\begin{aligned}
S_g^{(1)}[\phi] &\= -\tfrac{\im}{2}\,\Tr\,\tr\,\ln\,\bigl\langle\bar\psi(y)\,\psi(x)\bigr\rangle
+\tfrac{\im}{2}\,\Tr\,\tr\,\ln\,\bigl\{ -\bar{\slashed{\pa}}\Box^{-1}(y{-}x)\,f_g^{-2}(x) \bigr\} \\
&\= -\tfrac{\im}{2}\,\Tr\,\tr\,\ln\,\Bigl\{ \delta^{(4)}(y{-}x)\ +\
2\tfrac{g}{\mu^2}\,\bar{\slashed{\pa}}\Box^{-1}(y{-}x)\,f_g(x)\,\phi\,\slashed{\pa}\phi^*(x) \\
&\qquad\qquad\qquad\ +\ 
4\tfrac{g^2}{\mu^4}\!\int\!\diff^4z\ \bar{\slashed{\pa}}\Box^{-1}(y{-}z)\,f_g(z)\,\phi\,\slashed{\pa}\phi^*(z)\,
\bar{\slashed{\pa}}\Box^{-1}(z{-}x)\,f_g(x)\,\phi\,\slashed{\pa}\phi^*(x) + O(g^3) \Bigr\} \\
&\= (4-\tfrac122^2)\tfrac{g^2}{\mu^4}\!\int\!\diff^4y\!\int\!\diff^4z\ 
[\bar{\slashed{\pa}}\Box^{-1}(y{-}z)\,\phi\,\slashed{\pa}\phi^*(z)\,
\bar{\slashed{\pa}}\Box^{-1}(z{-}y)\,\phi\,\slashed{\pa}\phi^*(y)] + O(g^3)\ .
\end{aligned}
\end{equation}
We learn that the leading contributions to all three terms in~\eqref{detmatch} are proportional to
\begin{equation}
I \= \tfrac{g^2}{\mu^4}\!\int\!\diff^4x\!\int\!\diff^4y\ 
[\phi\,\slashed{\pa}\phi^*(x)\,\bar{\slashed{\pa}}\Box^{-1}(x{-}y)\,
\phi\,\slashed{\pa}\phi^*(y)\,\bar{\slashed{\pa}}\Box^{-1}(y{-}x)]\ ,
\end{equation}
so that the determinant-matching condition becomes
\begin{equation}
4\,I\ -\ 2\,I \= 2\,I\ ,
\end{equation}
which is obviously correct.\footnote{
The integral~$I$ is divergent and requires regularization. 
We assume that  this has been been done in an appropriate way.
The perturbative non-renormalizability of the nonlinear sigma models is not of concern. }
We remark that the classical Nicolai map~$T_g^{(0)}$
is not sufficient to saturate this condition. 
It is here that the leading quantum correction~ $T_g^{(1)}$ first makes a difference.

\section{Classicalizing the map}

\noindent
A supersymmetric nonlinear sigma model with a hermitian symmetric target space
can be reformulated as a gauged sigma model. To pass to this description,
one employs a Hubbard--Stratonovich transformation, which resolves the four-fermi interaction
in favor of a coupling to an (auxiliary) vector field. 
For $\C P^N$ models this was demonstrated in~\cite{CLR}, directly as well as via 
the superfield formulation of the gauged sigma model.
Without the fermion self-interaction, the Nicolai map 
(in the nonlocal bosonic theory now including the gauge field) 
may be classical, {\it i.e.\/} given entirely by fermionic tree diagrams.

Let us see how this comes about in the $\C P^1$ model.
The superfield formulation starts from the enhanced action~\cite{HN1,HN2},\footnote{
See~\cite{review} for a more general review on nonlinear realizations and hidden local symmetries.}
\begin{equation} \label{gaction}
\ensuremath{\mathring{\widetilde{S}}}_{\textrm{\tiny SUSY}} \ = 
\int\!\diff^4x \ \ensuremath{\mathring{\widetilde\Lagr}} \qquad\textrm{with}\qquad
\ensuremath{\mathring{\widetilde\Lagr}} \= 
\int\!\diff^2\theta\,\diff^2\bar\theta\ \Bigl\{ 
\ep^{\lambda V}\,\bigl(\tfrac{1}{\lambda^2}+\Phi^\dagger\Phi)\;-\;
\tfrac{1}{\lambda}\,V\;-\;\tfrac{1}{\lambda^2} \Bigr\}\ ,
\end{equation}
and corresponding expressions for $\ensuremath{\mathring{\widetilde\Delt}}$,
where an auxiliary real vector superfield~$V$ with components 
$(C,L,A_m,\lambda,\chi,D)$ has been introduced, and we defined the (only) dimensionful coupling
\begin{equation}
\lambda^2\ :=\ \frac{g}{\mu^2}
\qquad\textrm{so that}\qquad
f_\lambda = (1+\lambda^2\phi^*\phi)^{-1}\ .
\end{equation}
A complexified local U(1) gauge invariance has already been fixed completely,
but not in a Wess--Zumino gauge.
Since $\lambda$ has mass dimension~$-1$, the superfield~$V$ has mass dimension~$1$,
which means that the dimensions of its component fields are shifted by~$1$ from their canonical values.
In particular, the auxiliary vector field~$A_m$ is of dimension mass-squared.
Eliminating~$V$ by its algebraic equation of motion,
\begin{equation}
\ep^{\lambda V} \= \bigl(1+\lambda^2\,\Phi^\dagger\Phi\bigr)^{-1}
\qquad\Leftrightarrow\qquad
V \= -\tfrac{1}{\lambda}\,\log\bigl((1+\lambda^2\,\Phi^\dagger\Phi\bigr) 
\end{equation}
brings back the original action~\eqref{Kaction}.

We want to keep the auxiliary vector~$A_m$ but eliminate all other components~$(C,L,\lambda,\chi,D)$
as well as~$F$, in order to arrive at an effective (still local) bosonic theory for $(\phi,\psi,A_m)$.
With some algebra, this yields the enhanced lagrangian
\begin{equation}\label{Lagtilde}
\begin{aligned}
\widetilde\Lagr &\= 
-f_\lambda^2\,\partial_m \phi\,\partial^m \phi^* 
-\tfrac{\im}{2}f_\lambda^2 \bigl(\psi\,\slashed{\pa} \bar\psi+\bar\psi\,\bar{\slashed{\pa}} \psi\bigr)
+\tfrac{\im}{2}\lambda^2 f_\lambda^3\,\psi(\phi\,\slashed{\pa}\phi^*{-}\phi^*\bar{\slashed{\pa}}\phi)\bar\psi
+\tfrac14\lambda^2 f_\lambda^2 (\phi\,\pa_m\phi^*-\phi^*\pa_m\phi)^2 \\[4pt] 
&\qquad\! -\tfrac14\,A_m A^m 
+\tfrac{\im}{2}\lambda\,f_\lambda\,(\phi\,\pa_m\phi^*{-}\phi^*\pa_m\phi)\,A^m
+\tfrac12\lambda\,f_\lambda^2\,\psi\,\slashed{A}\,\bar{\psi} 
\end{aligned}
\end{equation}
and the penultimate component
\begin{equation}\label{Mtilde}
\widetilde\Delt \= -\im\sqrt{2}f_\lambda^2\,\slashed{\pa}\phi\,\bar\psi
+ \im\sqrt{2} \lambda^2\,f_\lambda^2\,\phi\,(\phi\,\slashed{\pa}\phi^*{-}\phi^*\slashed{\pa}\phi)\,\bar\psi
-\sqrt{2} \lambda\,f_\lambda\,\phi\,\slashed{A}\,\bar\psi\ .
\end{equation}
The algebraic equation of motion for~$A_m$ is solved by
\begin{equation}
A_m \= \im\,\lambda\,f_\lambda\,(\phi\,\pa_m\phi^*-\phi^*\pa_m\phi)\ +\
\lambda\,f_\lambda^2\,\psi\,\sigma_m\bar\psi\ .
\end{equation}
Inserting this back into~\eqref{Lagtilde} and~\eqref{Mtilde} and employing the Fierz identity
\begin{equation}
(\bar\psi \bar\psi)\,(\psi \psi) \= - \tfrac12(\psi \sigma_m \bar\psi)\,(\psi \sigma^m \bar\psi)
\end{equation}
indeed reproduces
\eqref{LagrCP1} and~\eqref{MCP1}, respectively.

The construction of the (enhanced) flow operator proceeds according to recipe,
\begin{equation} \label{flowenhanced}
\widetilde{R}_\lambda \=  -\tfrac{\im}{4}\int\!\diff^4{y}\!\int\!\diff^4{x}\
\bigl\langle\!\bigl\langle\pa_\lambda\ensuremath{\mathring{\widetilde\Delt}}_\alpha(y)\ 
\bigl\{ \delta^\alpha\phi(x)\,\tfrac{\delta}{\delta\phi(x)}\,+\; 
\delta^\alpha\!A_m(x)\,\tfrac{\delta}{\delta A_m(x)} \bigr\} \bigr\rangle\!\bigr\rangle
\end{equation}
where the double bracket $\langle\!\langle\ldots\rangle\!\rangle$ now stands for the functional average over
$(C,L,\lambda,\chi,D,F)$ and, of course, $\psi$. 
Contrary to the situation in Section~2, $\ensuremath{\mathring{\widetilde\Delt}}$ is now nonlinear in
the auxiliary fields, and hence we cannot eliminate them in~\eqref{flowenhanced} by inserting their
algebraic solutions back into the functional integral 
and writing a single bracket for the $\psi$~average. 
Because the proper elimination of all auxiliary fields in~\eqref{flowenhanced} is somewhat involved
we refrain from giving the details and just expose the result to order~$\lambda^2$.
The flow operator takes the form
\begin{equation}
\begin{aligned}
\widetilde{R}_\lambda &\= 
-\im\int\!\diff^4x\ \pa_m(\phi^*\phi)(x)\,\tfrac{\delta}{\delta A_m(x)} 
\ -\ \im\int\!\diff^4y\!\int\!\diff^4x\ \phi\,A(y)\cdot\pa\Box^{-1}(y{-}x)\,\tfrac{\delta}{\delta\phi(x)} \\
&\qquad\!+\lambda\int\!\diff^4y\!\int\!\diff^4x\ 
[\phi\,\slashed{A}(y)\,\bar{\slashed{\pa}}\Box^{-1}(y{-}x)\,
(\slashed{\pa}\phi^*(x)\,\bar{\sigma}_m + \phi^*(x)\,\pa_m)]
\,\tfrac{\delta}{\delta A_m(x)} \\
&\qquad\!-\lambda\int\!\diff^4y\!\int\!\diff^4x\ 
\phi\,(\phi\,\pa\phi^*{+}3\,\phi^*\pa\phi)(y)\cdot\pa\Box^{-1}(y{-}x)\,\tfrac{\delta}{\delta\phi(x)} \\
&\qquad\!-\tfrac12\lambda\int\!\diff^4z\!\int\!\diff^4y\!\int\!\diff^4x\
[\phi\,\slashed{A}(z)\,\bar{\slashed{\pa}}\Box^{-1}(z{-}y)\,\slashed{A}(y)\,\bar{\slashed{\pa}}\Box^{-1}(y{-}x)]
\,\tfrac{\delta}{\delta\phi(x)}
\ +\ O(\lambda^2)\ .
\end{aligned}
\end{equation}
With it, the (enhanced) Nicolai map becomes
\begin{align} \label{Nicenhanced}
\bigl(\widetilde{T}_\lambda\phi\bigr)(x) &\= \phi(x)\ +\ 
\im\lambda\int\!\diff^4y\ \phi\,A(y)\cdot\,\pa\Box^{-1}(y{-}x) \ +\ 
\lambda^2\int\!\diff^4y\ \phi^*\phi\,\pa\phi(y)\cdot\pa\Box^{-1}(y{-}x) \nn\\
&\qquad\!- \tfrac14\lambda^2\int\!\diff^4z\!\int\!\diff^4y\
[\phi\,\slashed{A}(z)\,\bar{\slashed{\pa}}\Box^{-1}(z{-}y)\,\sigma^n \bar{\slashed{A}}(y)\,\pa_n\Box^{-1}(y{-}x)]
\ +\ O(\lambda^3)\ ,\\[4pt]
\bigl(\widetilde{T}_\lambda A_m\bigr)(x) &\=  A_m(x)\ +\ 
\im\lambda\,\pa_m(\phi^*\phi)(x)\ -\ 
\tfrac12\lambda^2\int\!\diff^4y\
[\phi\,\slashed{A}(y)\,\bar{\slashed{\pa}}\Box^{-1}(y{-}x)\,\slashed{\pa}\phi^*(x)\,\bar{\sigma}_m]
\ +\ O(\lambda^3)\ .\nn
\end{align}
In the perturbative expansion, only tree diagrams will appear and no fermion loops. 
We can represent our result~\eqref{Nicenhanced} graphically,
\renewcommand\scale{.80}
\begin{align} \label{moregraphs}
\widetilde{T}_\lambda\phi &\=
\scalebox{\scale}{
\begin{tikzpicture}[baseline={([yshift=-1ex]current bounding box.center)},thick]
\node at (0,0)[circle,fill,inner sep = 2pt] {};
\draw[gluon](1,0)--(0,0);
\end{tikzpicture}
}
\ +\
\lambda \
\scalebox{\scale}{
\begin{tikzpicture}[baseline={([yshift=-1ex]current bounding box.center)},thick]
\node at (0,0)[circle,fill,inner sep = 2pt] {};
\draw (1,0)--(0,0);
\draw[gluon] (1+0.80, 0.60)--(1,0);
\draw[scalar] (1+0.80,-0.60)--(1,0);
\end{tikzpicture}
}
\ +\
\lambda^2\
\left(
    \scalebox{\scale}{
    \begin{tikzpicture}[baseline={([yshift=-1ex]current bounding box.center)},thick]
    \node at (0,0)[circle,fill,inner sep = 2pt] {};
    \draw (1,0)--(0,0);
    \draw[gluon] (1+0.75, 0.75)--(1,0);
    \draw[gluon] (1+1,0)--(1,0);
    \draw[gluon] (1+0.75,-0.75)--(1,0);
    \end{tikzpicture}
    }
     \ +\
    \scalebox{\scale}{
    \begin{tikzpicture}[baseline={([yshift=-1ex]current bounding box.center)},thick]
    \node at (-0.25,0)[circle,fill,inner sep = 2pt] {};
    \draw (2,0)--(-0.25,0);
    \draw[gluon] (2+0.80, 0.60)--(2,0);
    \draw[scalar] (2+0.80,-0.60)--(2,0);
    \draw[scalar] (1+0.0,-0.8)--(1,0);
    \phantom{ \draw[gluon] (1-0.5,0.8)--(1,0); }
    \end{tikzpicture}
    }
\right)
\ \quad + \ O(\lambda^3)\ ,
\\[4pt]
\widetilde{T}_\lambda A &\=
\scalebox{\scale}{
\begin{tikzpicture}[baseline={([yshift=-1ex]current bounding box.center)},thick]
\node at (0,0)[circle,fill,inner sep = 2pt] {};
\draw[scalar](1,0)--(0,0);
\end{tikzpicture}
}
\ +\
\lambda \
\scalebox{\scale}{
\begin{tikzpicture}[baseline={([yshift=-1ex]current bounding box.center)},thick]
\node at (0,0)[circle,fill,inner sep = 2pt] {};
\draw[gluon] (0.80, 0.60)--(0,0);
\draw[gluon] (0.80,-0.60)--(0,0);
\end{tikzpicture}
}
\qquad
\ +\
\lambda^2\ \quad
\scalebox{\scale}{
\begin{tikzpicture}[baseline={([yshift=-1ex]current bounding box.center)},thick]
\node at (0,0)[circle,fill,inner sep = 2pt] {};
\draw (1,0)--(0,0);
\draw[gluon] (0,-0.8)--(0,0);
\draw[gluon] (1+0.80, 0.60)--(1,0);
\draw[scalar] (1+0.80,-0.60)--(1,0);
\end{tikzpicture}
}
\ \quad +\ O(\lambda^3) \ ,
\end{align}
where the dashed lines represent insertions of the auxiliary field~$A$.
Compared to the classical map of Section~3, additional vertices appear:
Not only $(\phi,\phi^*)$ but also $A$ legs are attached to the branches of the trees,
but at most one at a given vertex. Because the $A$ propagator is ultra-local,
further averaging over~$A$ will connect vertices along a tree generating a four-fermi interaction,
thus reproducing the fermion loops of the quantum map of Section~4.
In other words, the enhanced Nicolai map resolves the loop decorations of the previous map,
effectively classicalizing it. The prize to pay is a more complex structure.

It is instructive to check the free-action condition. One easily obtains
\begin{equation}
\begin{aligned}
\smallint \widetilde{T}_\lambda\phi\,\Box\phi^* &\=
\smallint \phi\,\Box\phi^* + \im\,\lambda\smallint \phi\,A{\cdot}\pa\phi^*
+\lambda^2\smallint \phi^*\phi\,\pa\phi{\cdot}\pa\phi^*
-\tfrac14\lambda^2\smallint\!\!\smallint 
[\phi\,\slashed{A}\,\bar{\slashed{\pa}}\Box^{-1}\sigma^n \bar{\slashed{A}}\,\pa_n\phi^*]
+\ O(\lambda^3)\ ,\\[4pt]
-\tfrac14\smallint(\widetilde{T}_\lambda A)^2 &\=
-\tfrac14\smallint A^2 - \tfrac{\im}{2}\lambda\smallint A{\cdot}\pa(\phi^*\phi) 
+\tfrac14\lambda^2\smallint \pa(\phi^*\phi){\cdot}\pa(\phi^*\phi) 
+\tfrac14\lambda^2\smallint\!\!\smallint 
[\phi\,\slashed{A}\,\bar{\slashed{\pa}}\Box^{-1}\slashed{\pa}\phi^*\,\bar{\sigma}^m A_m]
+\ O(\lambda^3)\ .
\end{aligned}
\end{equation}
The sum nontrivally combines into
\begin{equation}
S_\lambda^{(0)}[\phi,A] \ =
\int\!\diff^4x\ \Bigl\{
(2\lambda^2\phi^*\phi{-}1)\,\pa\phi\cdot\pa\phi^*
-\tfrac14\,A^2
+\tfrac{\im}{2}\lambda\,(\phi\,\pa\phi^*{-}\phi^*\pa\phi)\,{\cdot}A
+\tfrac14\lambda^2 (\phi\,\pa\phi^*{-}\phi^*\pa\phi)^2
+ O(\lambda^3)\Bigr\}
\end{equation}
confirming our expressions~\eqref{Nicenhanced}.
A different but trivial solution of the free-action condition is ($g{=}\lambda^2\mu^2$)
\begin{equation}
\bigl(\widetilde{T}_\lambda\phi\bigr)(x) \= \bigl(T_{\lambda^2\mu^2}^{(0)}\phi\bigr)(x)
\qquad\textrm{and}\qquad
\bigl(\widetilde{T}_\lambda A_m\bigr)(x) \= A_m - \im\lambda\,f_\lambda\,(\phi\,\pa_m\phi^*{-}\phi^*\pa_m\phi)
\end{equation}
simply extending the previous classical Nicolai map~\eqref{Texplicit} 
by an almost trivial local map for the auxiliary vector~$A$.
However, this map cannot capture the fermion self-interaction and will not
fulfil the (refined) determinant-matching condition.
It again demonstrates the non-uniqueness of the free-action solutions.

\subsection*{Acknowledgments}
We thank Lorenzo Casarin for discussions.

\newpage


\end{document}